\documentclass[12pt,a4paper]{article}

\usepackage[cp1251]{inputenc}
\usepackage{amssymb} \usepackage{amsmath}
\usepackage{graphics}
\usepackage{xspace} %
\usepackage{float}
\newcommand\LB[1]{\label{#1}} 
\newcommand\BE[2]{\begin{#1} #2 \end{#1}}
\newcommand\ARR[2]{\BE{array}{{#1} #2}}
\newcommand\EQ[2]{\BE{equation}{\LB{#1} #2}}
\newcommand\EQA[3]{\EQ{#1}{\ARR{#2}{#3}}}
 
 \newcommand\EQn[1]{\BE{equation*}{ #1}}

\newcommand\EQAn[2]{\EQn{\ARR{#1}{#2}}}
\def\f{\varphi}

\newcommand\bg{{\mathbf g}}
\newcommand\bv{{\mathbf v}}

\newcommand\wrt{with respect to \xspace}
\newcommand{\atan}{\operatorname{atan}}

\newcommand \lm{\lambda}
\renewcommand\th{\theta}

\newcommand{\bH}{\mathbf{H}}

\renewcommand\d{\partial}

\title{
{\bf
Instantaneous Pole Velocity and Global Models
 }}

\author{
{S.A.Ivanov \thanks{Marine Geomagnetic Investigation Laboratory,
St.Petersburg Branch of Pushkov Institute of Terrestrial Magnetism, Ionosphere and Radio Wave
Propagation, 199164, St.Petersburg, Mendellevskaya, 1, Russia,. Corresponding author
{\tt sergei.a.ivanov@mail.ru}}
}
\and
{  S.A.Merkuryev
\thanks{Marine Geomagnetic Investigation Laboratory,
St.Petersburg Branch of Pushkov Institute of Terrestrial Magnetism, Ionosphere and Radio Wave
Propagation;
Saint Petersburg State University, Institute of Earth Sciences,Universitetskaya nab.,7-9, St. Petersburg 199034, Russia
}
 }
 \and
 {I. M. Demina
 \thanks{Marine Geomagnetic Investigation Laboratory,
St.Petersburg Branch of Pushkov Institute of Terrestrial Magnetism, Ionosphere and Radio Wave
Propagation, 199164, St.Petersburg, Mendellevskaya, 1
 }
  }
  }

\date{}
\begin{document}
\maketitle


\begin{abstract}
A new approach to estimating the velocity of the pole moving is considered.
The method uses the model-calculated spatial distribution of the (vector) horizontal component \textbf{H} for a given year and its change
relative to the closest epochs. The velocity equation is obtained from the expression for the time dependence of the position
of the pole  $\bH(\lm(t),\f(t),t)=0$. Here $\lm(t)$ and $\f(t)$
are the geographic coordinates of the pole at time $t$. The velocity between epochs can be found using the Hermite spline, which gives
a smooth line that keeps the velocity vector in each epoch. We use IGRF and COV-OBSx2 models to find the horizontal component.
\end{abstract}

\section {Introduction}

The magnetic pole is a conditional wandering point on the surface of the northern (southern) hemisphere
of the Earth, where the geomagnetic field is directed vertically, i.e., perpendicular to the ellipsoid,
and its horizontal component is zero. This property was used when searching for the position of this point using analytical models.
Studies show that the position of the Earth's magnetic pole is constantly changing, and the rate of this change is not constant in time
and differs for the North and South magnetic poles.

The real movement of the magnetic pole is an integral characteristic. Its secular trend reflects deep processes in the Earth's core,
which determine both global  and regional anomalies of the geomagnetic field \cite{1,2}. In addition, the current position of the pole
is greatly influenced by the variable part of the geomagnetic field, especially on disturbed days. The fluctuations associated with this
can be tens of kilometers \cite{3}. The increased publication activity in recent years and interest in the shift
of the Earth's North Magnetic Pole (NMP) are due to the fact that over the past 20 years the rate of its shift has increased by more than
three and a half times. So, if in the mid-1990s the pole was moving at a speed of 15 km per year, now the speed has increased
to 55 km per year, \cite{4}. In almost all the literature devoted to the study of the  movement of magnetic poles,
global analytical models are used  to obtain the interval (average) velocity and
direction of movement between epochs \cite{5}. As a rule, these are IGRF \cite{6}, \emph{gufm }\cite{7}, and COV-OBSx2 \cite{8}.
These global models are built for successive epochs in increments of five years for IGRF, two and a half years for \emph{gufm,}
and two years for COV-OBSx2.

We propose a method for calculating both the magnitude and direction of the instantaneous velocity of the poles based on these models.
First, we test the algorithm on a dipole field whose pole moves with constant velocity. In this ase the error of the method depends
only on errors of numerical differentiation. Then we test the method on a reduced IGRF model. After this we apply the algorithm to full
IGRF and COV-OBSx2 models and discuss the results.

The calculation is performed for successive epochs, and interpolation between epochs is given using Hermite splines.
As a result of their application, we construct a smooth pole trajectory between epochs.It turns out that for the COV model,
in comparing to the IGRF model, this method gives more adequate results. One of the reason is that the interval between COV-models
is two years? not five one as in IGRF model.

\section{Equations of Instantaneous Velocities of the Poles}

From global models, one can obtain not only the position of the magnetic pole and the average speed of its movement,
but also an estimate of the speed and direction of its movement at a given time. This approach is used in \cite{9},
where the instantaneous velocity is estimated under the assumption that the direction of the pole movement
is known for the period of maximum acceleration of the NMP drift in 1989–2002. The authors show how a change in drift velocity
is related with a spatial and temporal change of the field projection on the direction of motion.

We propose the following approach to determine the instantaneous velocity. Let in the vicinity of the pole for a given epoch,
the spatial distribution of the  horizontal component $\textbf{H}$ of the geomagnetic field be known,
as well as the changes in $\textbf{H}$ over time at all points in this region. Then it is possible to obtain the value of the instantaneous velocity
and the direction of the magnetic pole movement by calculating the derivative of  $\textbf{H}$  with respect to time and coordinates.
Let us introduce the following notation. Suppose that the pole moves with time along the curve $(\lm(t), \f(t))$, where $\lm$ and $\f$
are the longitude and the latitude of the pole at time $t$. Then, by the definition of the pole, we obtain the equation for
the horizontal component of the field:
$$
\textbf{H}((\lm(t), \f(t),t)=0.
$$
In the $X$, $Y$ components of the field we find
\EQAn{rl}{
X(\lm(t), \f(t),t)=&0 \\
Y(\lm(t), \f(t),t))=&0
}
Differentiating these expressions with respect to $t$ as a superposition, we obtain
\EQAn{rl}{
\frac{dX}{ dt}=&\frac{\d X}{\d\lm} \frac{d\lm}{dt}+\frac{\d X}{\d\f} \frac{d\f}{dt}+\frac{\d X}{\d t}=0,\\
\frac{dY}{ dt}=&\frac{\d Y}{\d\lm} \frac{d\lm}{dt}+\frac{\d Y}{\d\f} \frac{d\f}{dt}+\frac{\d Y}{\d t}=0.
}
Considering the field and its changes at time $t$ as known from an analytical  model for a given epoch, we can write
\EQ 2{
A \bv=\bg,
}
where
\EQ 3{
A=
\begin{pmatrix}
\d X/\d\lm& \d X/\d\f       \\
\d Y/\d\lm & \d Y/\d\f
\end{pmatrix},
}
and
\EQ 4{
\bv=\binom{ d\lm/dt}{ d\f/dt},\  \bg=-\binom{\d X/\d t}{\d Y/\d t}.
}
Solving this system, we find the pole velocity at time $t$.

\section{Numerical implementation of the algorithm and  numerical differentiation}

As it seen from \eqref2, \eqref4 the velocity strongly depend on the time and spatial
derivatives  of the field. The patial derivatives can be found analytically since the Gauss coefficients are known, or
 or numerically because the field variations are  small (there are no high frquency terms).
Because the analytical field is given with time interval equal to several years,
we have to tested the algorithm \wrt time variations.

Let us test the algorithm on models. First, we take such a time-varying dipole field that its poles move at a constant veocity. This will allow not to use the formulas for instantaneous velocity for verification, since the average velocity coincides with the instantaneous one, and the average (interval) velocity can obtained from a comparison of the positions of the poles of two adjacent epochs. At the second step the algorithm is tested \wrt the reduced IGRF model.

\subsection{ Dipole poles }

It is known that the general case of a dipole field is produced by the potential
$$
U(\lm, \th)=Rg_{10}\cos\th +R[g_{11}\cos\lm + h_{11}\sin\lm ]\sin\th
$$
(we consider the field on the sphere or the radius $R$).

The horizontal components in the local geodetic coordinate system (NED)  are
\EQA{dU}{rl}{
B_x=\frac{\d U}{R\d \th}=                         &-g_{10}\sin\th +[ g_{11}\cos\lm + h_{11}\sin\lm]\cos\th\\
B_y=-\frac1{R\sin\th}\frac{\d U}{\d \lm}=& g_{11}\sin\lm - h_{11}\cos\lm.
}
From here let us find the known expressions for the poles of the dipole. Start with the longitude. From \eqref{dU} we see for poles
\EQ{th_p}{
\tan \lm=h_{11}/g_{11},
}
what gives two solutions corresponding two poles
\EQ{lp}{
\lm_p^-=\atan \frac{h_{11}}{g_{11}},\  \lm_p^+=\atan \frac{h_{11}}{g_{11}}+\pi.
}
From here setting $C=\sqrt{g_{11^2}+h_{11^2}}$ we obtain
$$
\cos^2\lm_p^\pm=\frac1{1+\tan^2\lm_p^\pm}=\frac {g^2_{11}}{C^2}, \
\sin^2\lm_p^\pm=\frac{\tan^2\lm_p^\pm}{1+\tan^2\lm_p^\pm}=\frac {h^2_{11^2}}{C^2},
$$
and
$$
\cos\lm_p^\pm=\pm\frac {g_{11}}C, \
\sin\lm_p^\pm=\pm\frac {h_{11}}C.
$$
Here the signs $\pm$ are  consistent in order that $\tan \lm_p^\pm$ satisfy  \eqref{th_p}.

From here
$$
g_{11}\cos\lm_p^\pm + h_{11}\sin\lm_p^\pm=\pm C
$$
and
\EQ s{
\tan \th_p^\pm=\pm\frac C{g_{10}},\ 0\le\th_p^\pm\le\pi.
}
Evidently, small positive value of $\th_p^\pm$ corresponds to the NMP.  Because  $g_{10}$ is negative ( IGRF), we take the sign $'-'$ in \eqref s. Therefore the NMP coordinates are
\EQ{p}{
\lm_p=\atan \frac{h_{11}}{g_{11}},\  \th_p=-\atan \frac{C}{g_{10}}.
}
For the SMP we have
\EQn{
\lm_p=\atan \frac{h_{11}}{g_{11}}+\pi,\  \th_p=\atan \frac{C}{g_{10}}.
}

\subsection{ Uniformly moving dipole poles and testing the algorithm on the dipole field}

It is possible to find  such time-dependence of the dipole Gauss coefficients that the pole moving with  constant velocity. Below
$(\lm_p(t),\th_p(t)$ is the North magnetic pole (NMP) position, $q$, $a$ and  $b$ are  numbers.

Case 1. Set
$$
g_{10}=q=Const, \  g_{11}=\cos at,\  h_{11}=\sin at.
$$
Then  by \eqref p
$$
\lm_p=at, \ \th_ p=-\atan1/q.
$$
The pole moves with constant velocity at fixed latatude.

Case 2. Set
$$
g_{10}=\cos at, \  g_{11}=+0 (\text{positive and very small}),\  h_{11}=\sin at.
$$
Then  by \eqref p
$$
\lm_p=\frac\pi2, \ \th_ p=-at.
$$
The pole moves with constant velocity at fixed longitude.

Case 3. Set
$$
g_{10}=1/\tan bt, \  g_{11}=\cos at ,\  h_{11}=\sin at.
$$
Then  by \eqref p
$$
\lm_p=at, \ \th_ p=-bt.
$$
The pole moves in a straight line (un the geographic coordinates) with constant velocity.

Collect all in a table

\begin{figure}[H]
\includegraphics{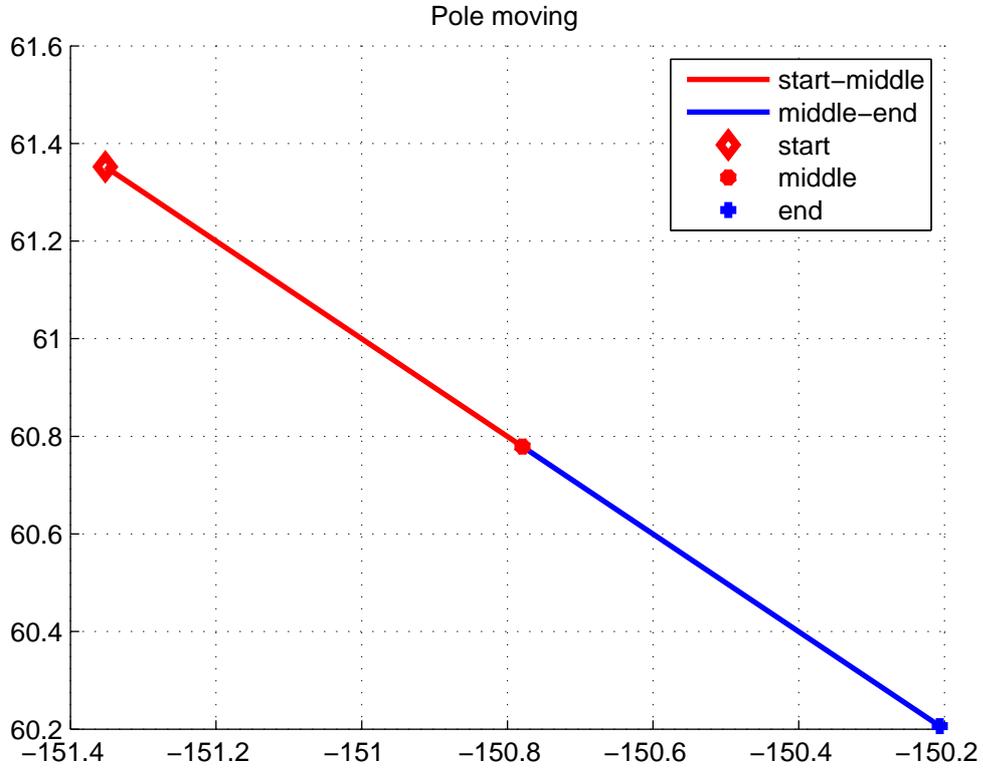}
\caption{\label{dipCvel} The pole location for a dipole field with constant pole velocity}
\end{figure}

\begin{tabular}{||c|c|c|c|c|c||} 
\hline
$g_{10}$&$g_{11}$&$h_{11}$  &$\lm_p$&$\th_p$\\
\hline
\hline
$q=Const$   &$\cos at$&$\sin at$  &$at$ &$-\atan1/q$\\
\hline
$\cos at$       &$+0$      &$\sin at$         &$\pi/2$ &$-at$\\
\hline
$-1/\tan bt$   &$\cos at$&$\sin at$         &$at$      &$-bt$\\
\hline
\hline
\end{tabular}
\vskip2mm
Figure \ref{dipCvel} show that the pole trajectory is a strait line in longitude-latitude coordinates.
Here we take the third case with $a=b=1$, the time points are 0.5, 0.51, and 0.52.   For these points we find the pole locations
by \eqref p.  The poles found by velocity  have a shift about $0.01^\circ$ \wrt the "true" pole becuse the numerical differentiation error. The point is that the magnetic fielf change is not linear in contrast to pole locations. 
Note that for the time step 0.1 the error in the latitude is $1^\circ$.

\subsection{ Testing the algorithms on a reduced IGRF}

\begin{figure}[H]
\includegraphics{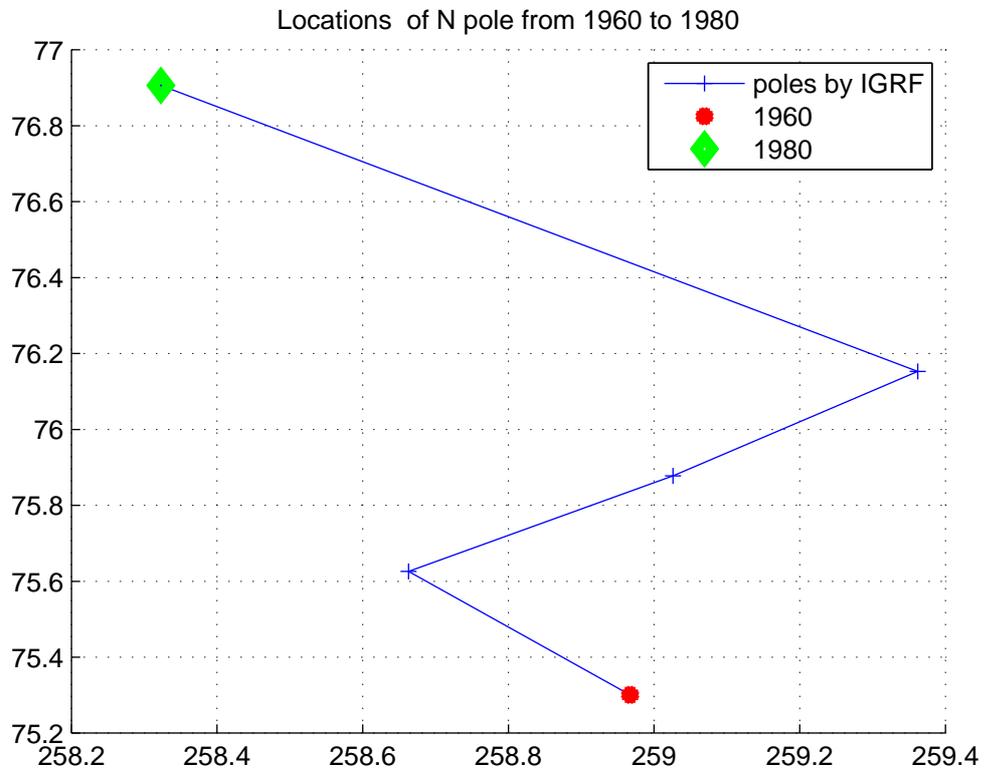}
\caption{\label{N60_80} The NMP by IGRF for 1960 - 1980 years. Maximal angle between adjacent interval  velocities is  33.6 degrees at 1975. }
\end{figure}

\begin{figure}[H]
\includegraphics{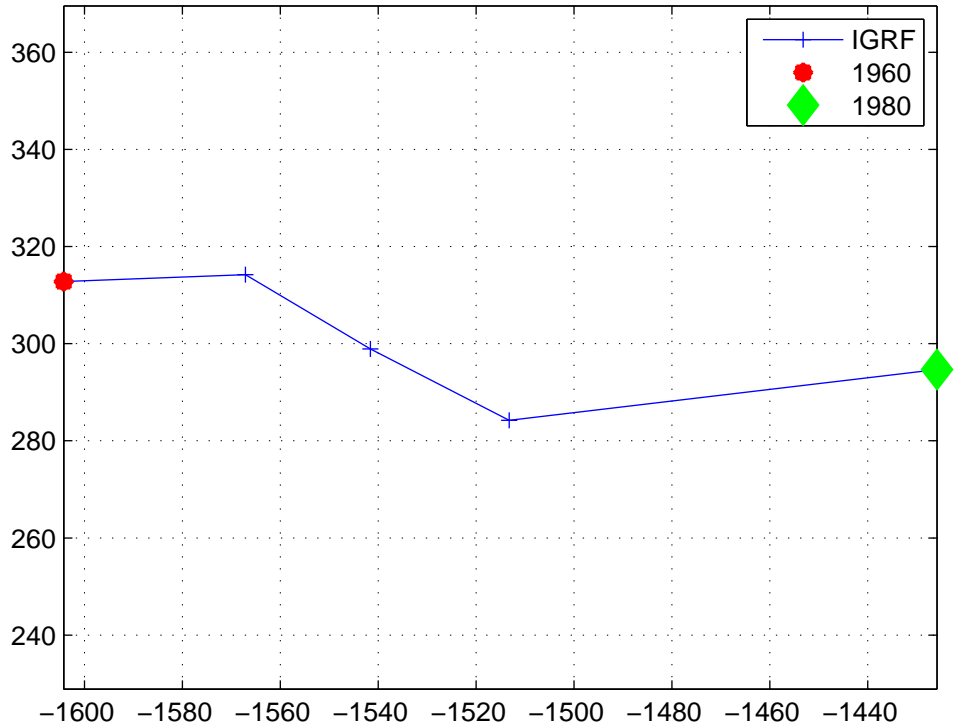}
\caption{\label{N60_80_Az} The NMP location by IGRF for 1960 - 1980 years in  the azimuthal equidistant projection
with the center in the Notrh geographic pole. Maximal angle between adjacent interval  velocities is  33.6 degrees at 1975. }
\end{figure}

Using the IGRF model to calculate the time derivative, we have field values only in increments of 5 years,
what can lead to numerical differentiation errors. This is especially clearly seen from Fig. \ref{N60_80},
where the segments of the trajectory of the NMP are shown for the periods of  the greatest jumps of the interval pole velocity.
The scales of the figure stress these jumps, nevertheless the angle between the  interval velocities for 1970-1975 and for 1975-1980
is  33.6 degrees.  To show (only here) the real trajectory of the pole, see it (in the azimuthal equidistant projection  on Fig.
 \ref{N60_80_Az}.
 The same is true for the SMP, the angle between the  interval velocities for 1940-1945 and for 1945-1950
  is 44.7 degrees, see Fig.\ref{S35_50}.

\begin{figure}[H]
\includegraphics{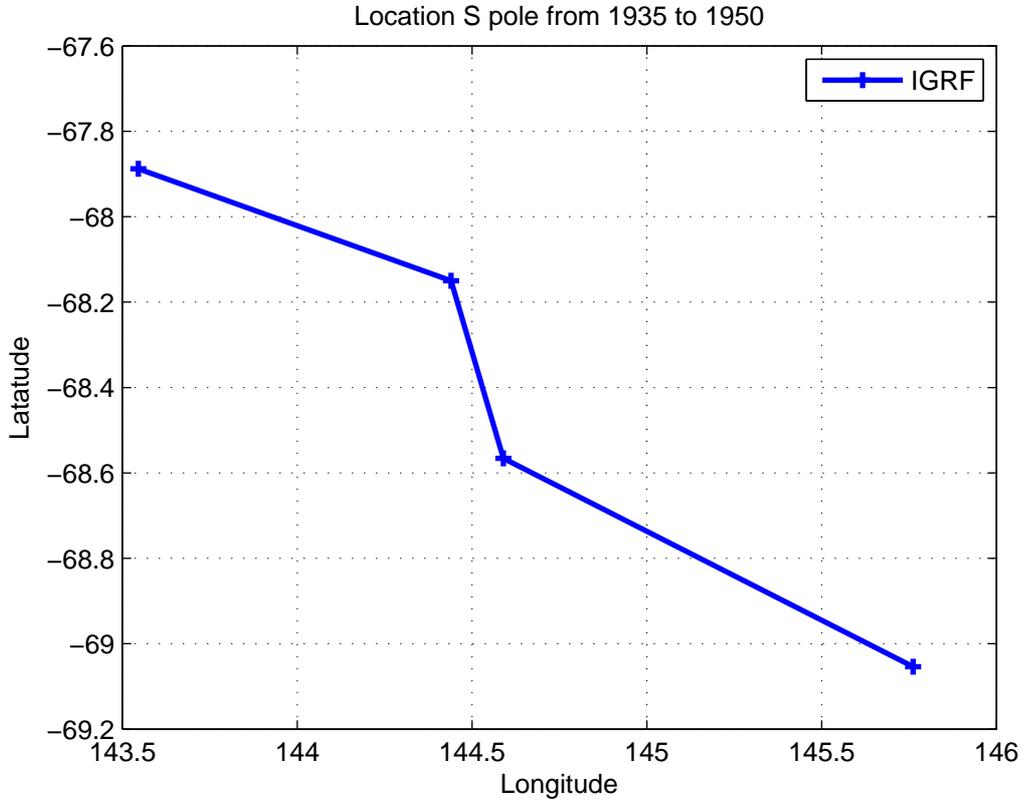}
\caption  {\LB{S35_50} The SMP location by  IGRF for 1960 - 1980 years. Maximal angle between adjacent interval velocities is  44.7 degrees. }
\end{figure}

We can consider  three variants for calculating the time derivative: using knowledge of the field
of the present and future epoch (we denote the speed calculated by this method by F–forward), knowledge of the previous and present
epoch (B, backward) and, finally, the previous and future epoch (S - symmetric). For smooth curves, the latter method gives an error
of an order smaller than the first two. Obviously, for the situations depicted in Fig, \ref{N60_80}, these options will give sharply
different values. We think there are no sense to use higher order approximation of time derivative here, say five-point one.

In the IGRF model the geomagnetic field is given as a finite sum of harmonics with the minimal period about 3500 km,
and the numerical differentiation in spatial variables is not hard.  We us the forward approximation with the step equal to  a quarter degree.

The algorithm was tested on models. According to a set of Gauss coefficients up to the octupole (that is, up to the 3rd order),
the distribution of the components $X$ and $Y$  was calculated, for example, for the IGRF epoch of 1925 and 1930 years.
The position of the pole was determined by the minimum $H$ (calculations were carried out on the surface of the sphere).
In accordance with formula \eqref3, the derivatives of the components $X$ and $Y$ at the pole point were calculated.
As a field for the next epoch, either a field obtained by a random perturbation of the Gauss coefficients of the present epoch was taken,
or a field determined by a set of coefficients corresponding to the next epoch, the IGRF of 1930. The time derivative of the components
 $X$ and $Y$ was calculated numerically, equal to  an increment divided by the time interval (forward scheme).  According to \eqref2-\eqref4,
 the instantaneous velocity was found in units of degrees/year.
Based on the found velocity we find the "forecasting pole". It is the the point
to which the pole will move from the previous epoch to the new one at the found velocity.
Further, according to the octupole model, the pole for the new epoch was calculated as the minimum $H$ for the new coefficients.
The new pole was compared with the forecasted pole.

In Fig. \ref{oct1925}  the result of this algorithm for the coefficients IGRF of the two epochs 1925 and 1930 (with differentiation F) is shown.
The deviation of the forecasted pole from the pole calculated for the same
epoch using the reduced IGRF model is hundredths of a degree, what corresponds to the  step of the pole determination.

\begin{figure}[H]
\includegraphics{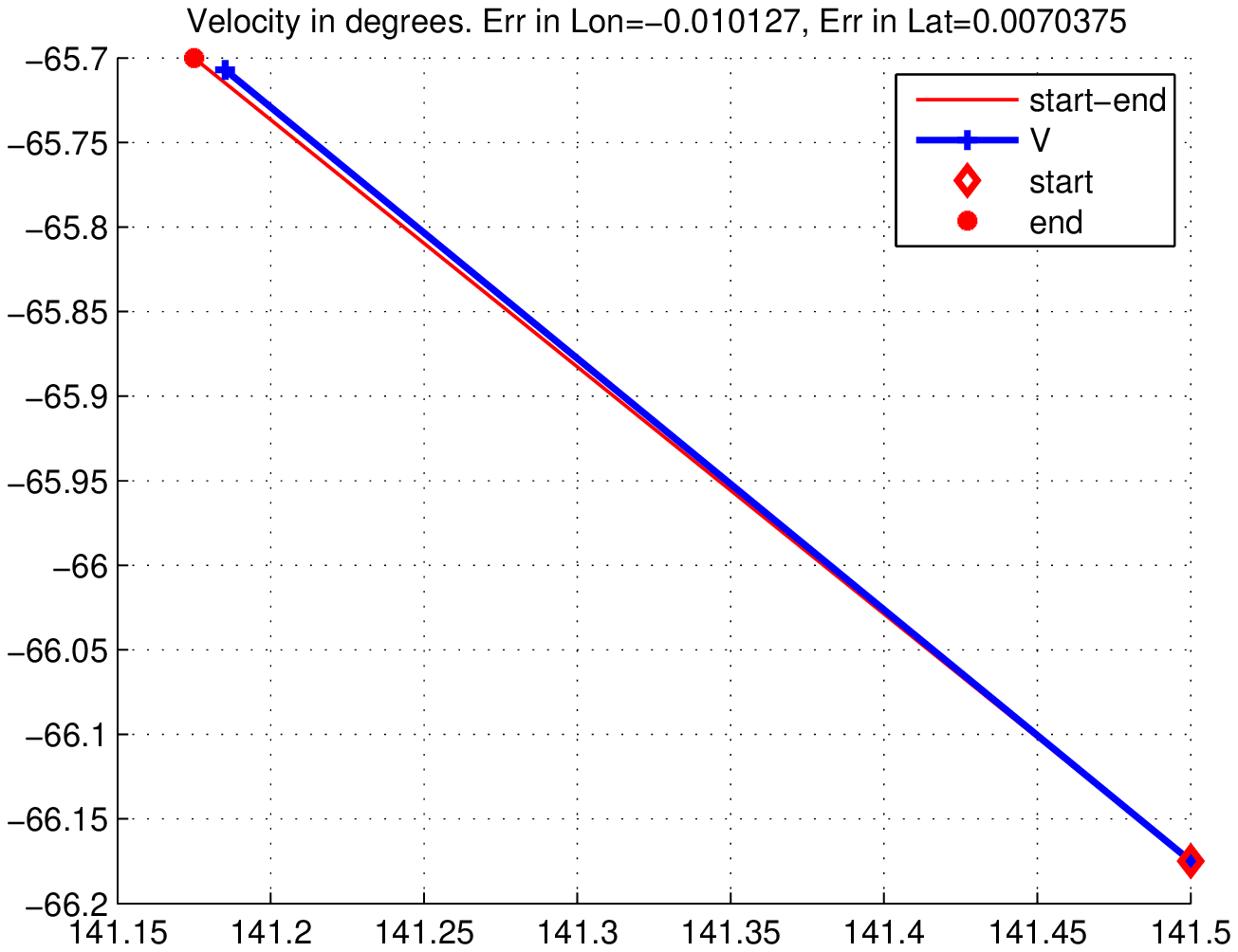}
\caption{\LB{oct1925} An example of the algorithm for the first nine IGRF coefficients of two epochs 1925 - 1930.
The red diamond shows the position of the SMP for the epoch of 1925 according to the IGRF octupole model, the circle  -
for the epoch of 1930, the blue cross shows the position of the pole in 1930, calculated by the instantaneous velocity. The deviation is 0.85 km. }
\end{figure}

\section{ Instantaneous   velocities and splines for the IGRF and COV models}

\subsection{ Instantaneous  velocities for the IGRF model}

\begin{figure}[H]
\includegraphics{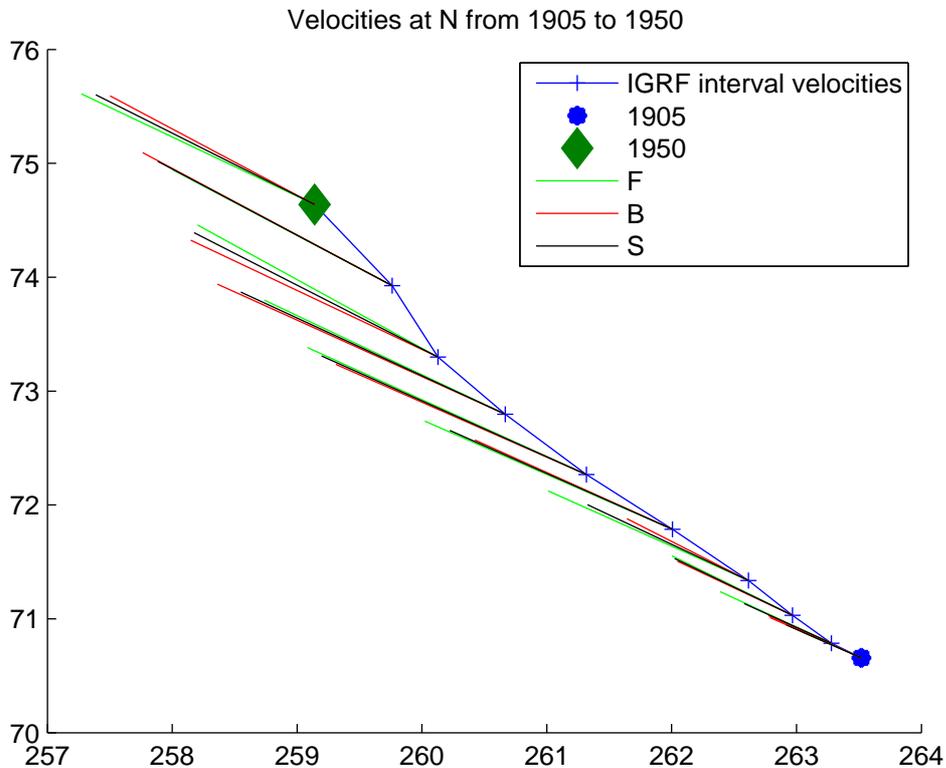}
 \caption{  \label{Fig4} Instant velocities for NMP.  The positions of the poles for the epochs 1905-1950
   are shown  with blue crosses. The lines show the instantaneous velocities calculated according to three variants of numerical differentiation. }
\end{figure}

First we show the results for the years  1905-1950, when the interval velocities change  slow: the maximal change
of the interval velocities is about $9^\circ$.
The positions of the NMP  for the epochs of 1905-1950 were calculated using the IGRF model, and are shown in Fig.\ref{Fig4}
with blue crosses. The lines show the instantaneous velocities calculated according to three variants of numerical differentiation.
For clarity, the speed values were increased by 5 times. This gives the position of the forecasted pole for the next epoch.

We see on the figure that all variants are close one to another but all have a systematic error \wrt the interval velocities.
We have not adequate explanations of this artefact. Partially this connected with the
large contribution of high order spherical harmonics in polar regions.
If we compare Fig.\ref{oct1925}  with the corresponding part of the Fig. \ref{Fig4}
we find that the instantaneous velocity  of the reduced octupol IGRF model gives a good approximation of the interval velocity this model
reduced up the third order. This is true for all epochs under considerations. Calculation gives a large difference between the horizontal field
of the full and the IGRF models for polar regions:

For the latitude $70^\circ$ and the longitude $260^\circ$ the components of the octupol model
are $ X=1438\, nT$,  $Y=417\,  nT$. For the full model we have $X=5476\,  nT$, $Y=3196\,  nT$. The root squared error is 4900 nT.
For the equatorial region the root squared error is 260 nT (the latitude $0^\circ$ and the longitude $260^\circ$).

Consider now  epochs with a large change of the adjacent  interval velocities. In this case, evidently,  the instant
velocity found by this method strongly depends on the choosen numerical time-differentiation, see  Fig. \ref Z.
Also we see that at the interval 1965-1970 the best approximation of the interval velocity gives the S-differentiaion
(equal to the average of the F- and B-variants).

\begin{figure}[H]
\includegraphics{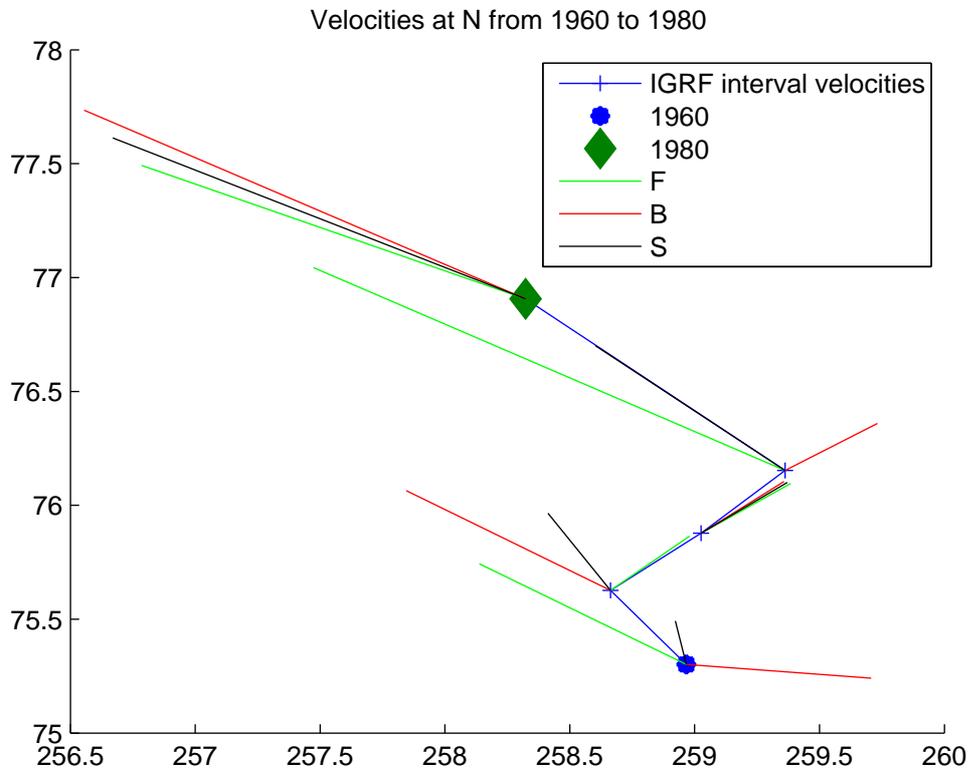}
\caption{\label{Z} The instantaneous velocities of   NMP by IGRF for 1960 - 1980 years. Maximal angle between adjacent interval
velocities is  33.6 degrees. }
\end{figure}

\subsection{ Hermite splines for the IGRF model}

Usually  Hermitian splines are use for functions. Since the trajectory is a moving on a 2dim surface, we construct the splines as follows.
For the IGRF model we have points (nodes) $(\lm_n,\f_n)$ of  magnetic poles in 5 years. At the nodes let us assign the derivatives
$\lm'_n$, $\f'_n$. Now we can construct   Hermitian splines  $\lm(t)$ and  $\f(t)$ for the longitude and the latitude. The moving
$(\lm(t),\f(t))$ we call  the spline for poles moving. Spline match the nodes and have the velocity $(\lm'_n,\f'_n)$ at the nodes. If we exclude time
we will have a curve. The tangent line at nodes  coincide with the direction of vectors $(\lm'_n,\f'_n)$.
We have four set of velocities: interval  ones and three variants of a instant velocities.

\begin{figure}[H]
\includegraphics{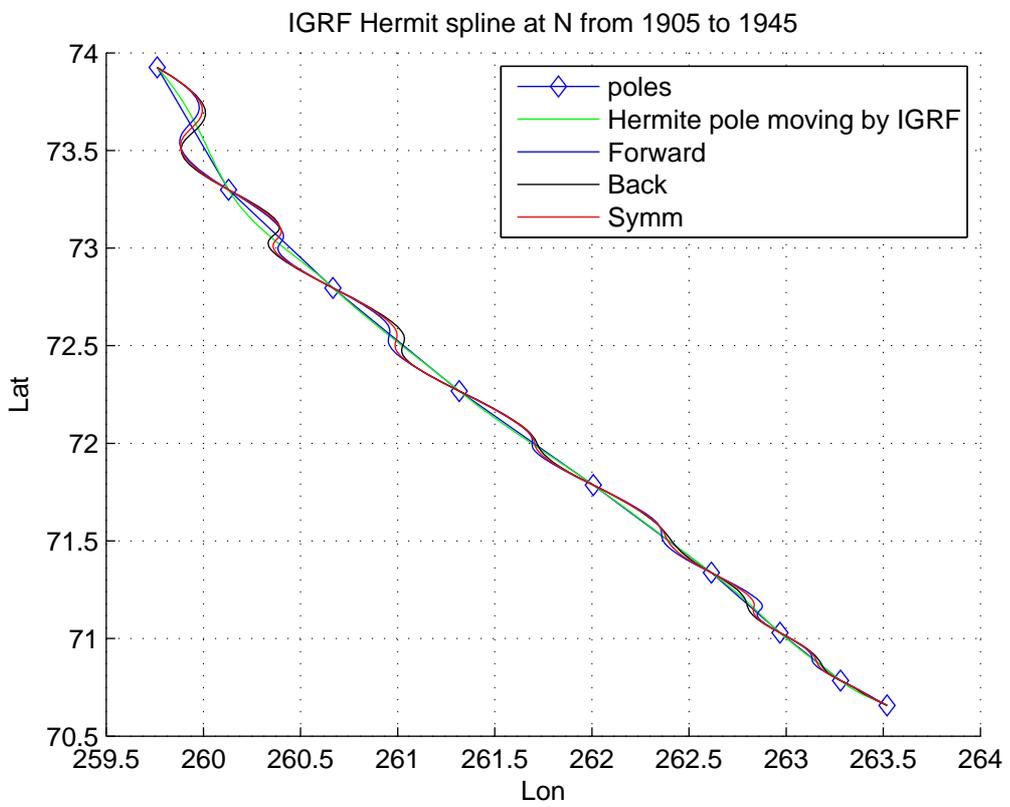}
\caption{\label{vIGRF_05_50}  Hermite splines  for SMP by the IGRF  model:  1905-1950.  }
\end{figure}

 Figure \ref{vIGRF_05_50} shows the results of calculation and the pole positions for the epochs of 1905-1940 according to the IGRF model.
 The green curve is a Hermite spline constructed from interval (average) velocities. The other three curves correspond
  three variants of the time derivative calculating. From Fig. \ref{vIGRF_05_50} it can be seen that
 the curve constructed from interval velocities differs significantly from the other curves,
 which have fluctuations with a period of 5 years, coinciding with the interval between epochs.

 We assume that this is an artifact associated rather not with numerical differentiation errors,
 but with long interval between models, with the method of constructing IGRF models. See also the discussion about trduced models above

\begin{figure}[H]
\includegraphics{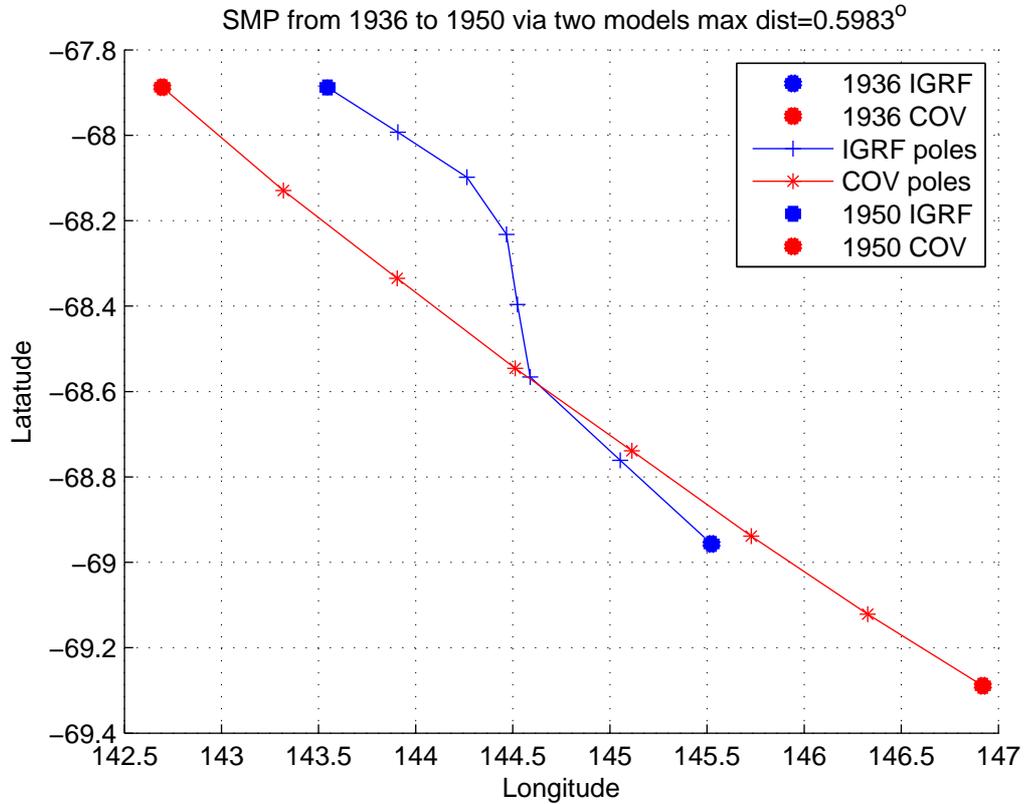}
\caption{\label{comp} SMP poles by the IGRF and COV  model:  1936-1950.  }
\end{figure}

 Note also that different analytical models of the geomagnetic field give different estimates of velocities \cite{9}.
 This can also be seen from the fact that the position of the poles of two models  may differ significantly.
 For illustration, Fig. \ref{comp}compares the pole positions of two models in the period 1936-1950. In the interval between the epochs,
 the poles of the IGRF model are taken from [https://www.ngdc.noaa.gov/geomag/data/poles/SP.xy ]. The poles of the COV
 model were obtained by the authors (the main contribution was made by Demina). The maximum distance between the positions
 of the south poles calculated from the global geomagnetic field models IGRF and COV for the same epoch is 0.60 degrees of the
 arc of the great circle, that is, about 60 km. At the same time, the direction of the interval velocity for the corresponding epochs may differ by $42^\circ$.
 Obviously, it is hard to expect high accuracy in calculating the time derivative in such a situation.

\subsection{ Instantaneous  velocities for the COV-OBSx2  model}

\begin{figure}[H]
\includegraphics{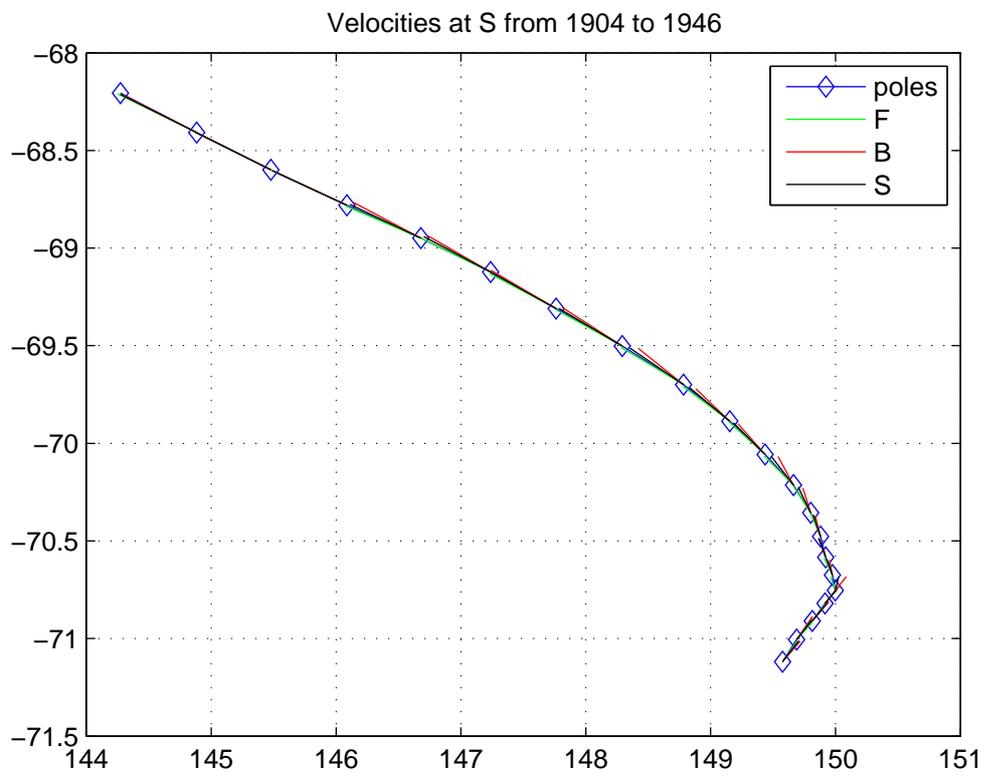}
\caption{\label{COV_vel_04_46} Instant velocities for SP by the COV model.  }
\end{figure}

\begin{figure}[H]
\includegraphics{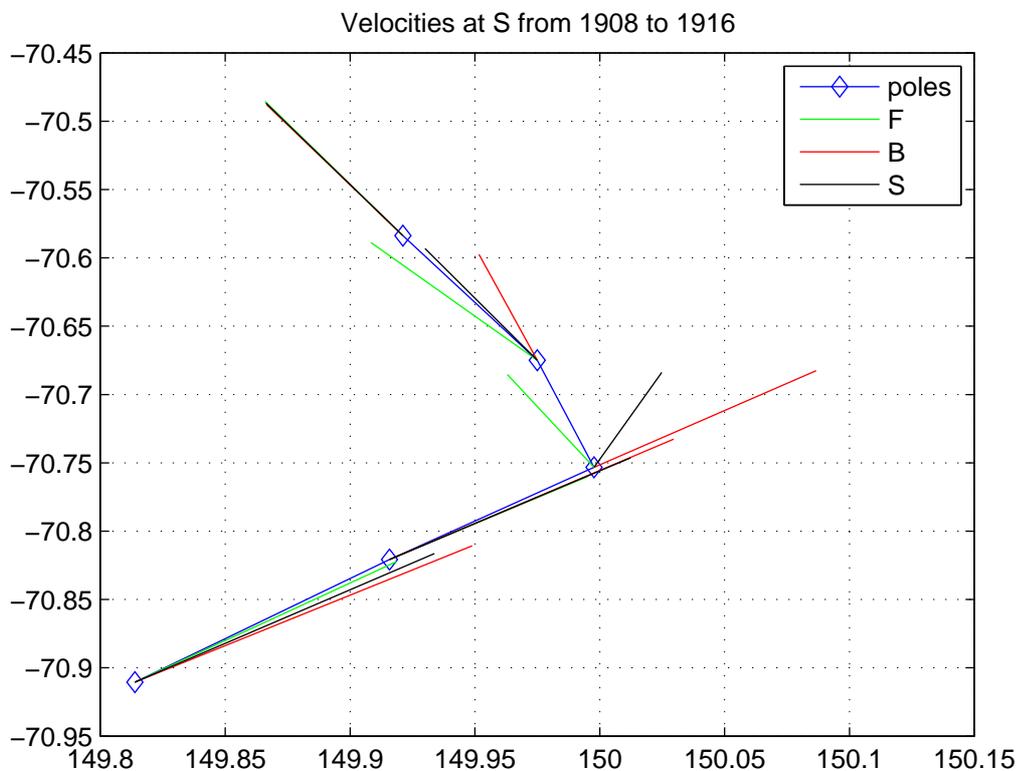}
\caption{\label{vCOV08_16} Instant velocities for SP by the COV model.  }
\end{figure}

Unlike the IGRF model, the COV-OBSx2 model created at intervals of two years. As can be seen from Fig. \ref{COV_vel_04_46},
 the trajectory of the SMP  calculated by this model is a smoother  trajectory \wrt the IGRF model almost where,
 what  affects the results of calculations of instantaneous velocities. Nevertheless in 1912 the interval velocities turns by $27^\circ$.
 The velocities at close epochs are shown on Fig.
\ref{vCOV08_16}. We see that the interval velocities are close to the F-instantaneous velocities. Note that the angle between
the B- and interval velocitis arrive $68^\circ$.

Note that the velocity are rotated from the arc of the trajectory of the poles movement.

\subsection{ Hermite splines for the COV-OBSx2  model}

\begin{figure}[H]
\includegraphics{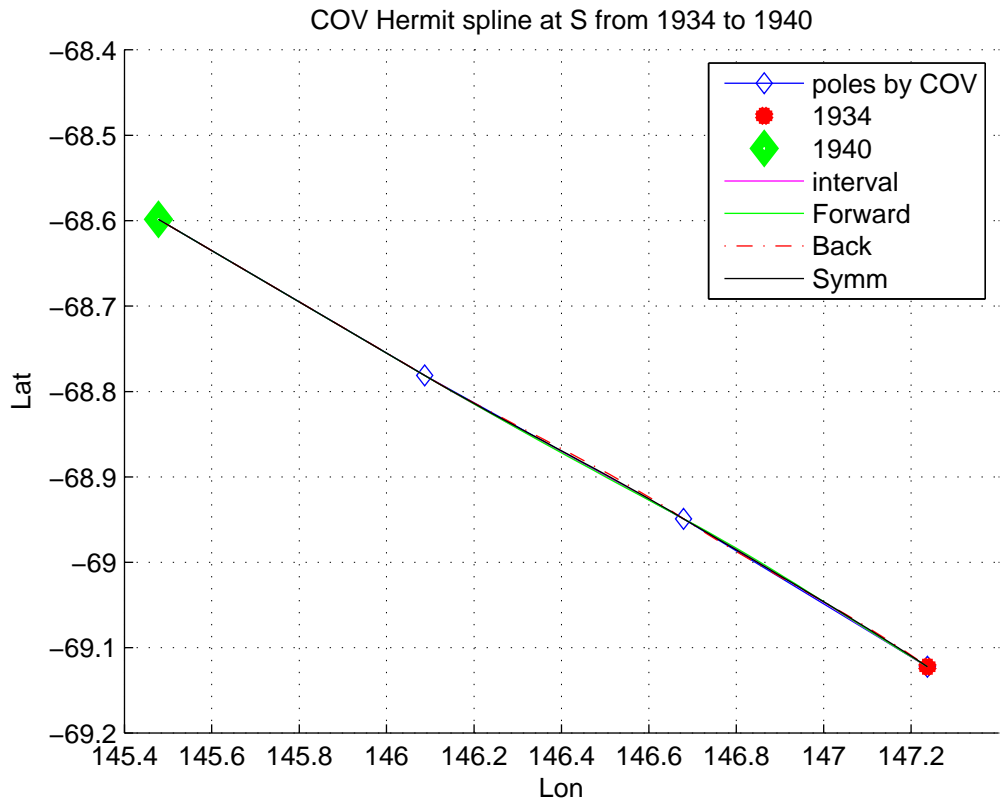}
\caption{\label{hCOV_34_40} Hermite splines  for SMP by the COV model.  A smooth part. }
\end{figure}

For the "smooth" parts of the trajectory all Hermit splines very close to the  trajectory. See, for example,   Fig. \ref{hCOV_34_40}.
The angles between the adjacent interval velocities is less than $3^\circ$.

\begin{figure}[H]
\includegraphics{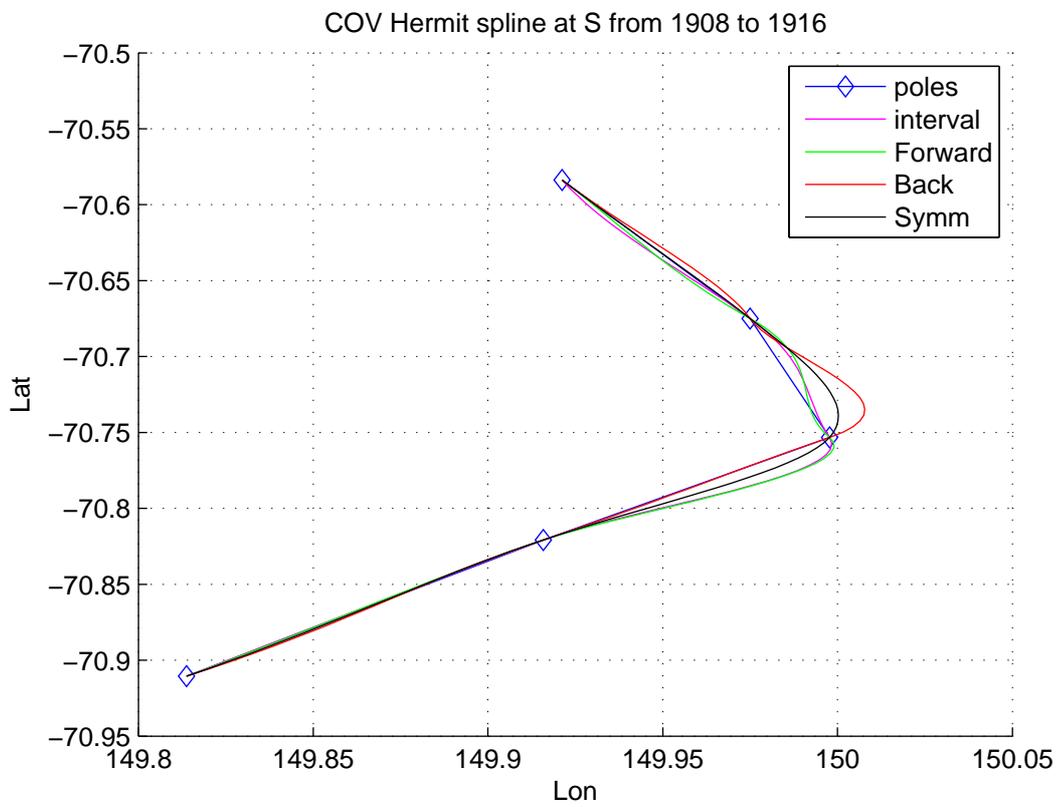}
\caption{\label{hCOV_08_16} Hermite splines  for SMP by the COV model.  }
\end{figure}

The hermite spline corresponding the 1908-1916 time interval is shown in Fig. \ref{hCOV_08_16}.
We conclude that the COV model gives more reliablr results

\newpage

\begin{thebibliography}{12}

\bibitem{1}	Demina, I.M., Nikitina, L.V. \& Farafonova, Y.G. Secular variations in the main geomagnetic field
within the scope of the dynamic model of field sources. \emph{Geomagn. Aeron.} 48, 542–550 (2008).
https://doi.org/10.1134/S0016793208040166

\bibitem{2}	 Merkuryev S.A.,  Demina I. M., Ivanov S.A. New data on the South magnetic pole location  in comparison with global models.
\emph{Proceedings "Geocosmos 2022}".

\bibitem{3}	Semakov N.N., Kovalev A. A., Pavlov A. F. Fedotova O. I. Where does the magnetic pole run? \emph{"First-hand Science"}
Vol. 68. No. 2. pp. 97-106. 2016

\bibitem{4}	Witze A.Earth's magnetic field is acting up // Nature.V.565.N.7738.P.143-144.2019.https://doi.org/
10.1038/d41586-019-00007-1
\bibitem{5}	Korte M.,Mandea M.Magnetic poles and dipole tilt variation over the past decades to millennia //
 Earth Planets and Space.V.60.N.9,P.937-948.2008.
\bibitem{6}	Alken,P.,Thebault,E.,Beggan,C.D.et al.International Geomagnetic Reference Field: the thirteenth generation /
/ Earth Planets Space.V.73.49.2021.https://doi.org/10.1186/s40623-020-01288-x
\bibitem{7}	Jonkers A.R.T.,Jackson A.,Murray A.Four centuries of geomagnetic data from historical records.//
 Rev.Geophys.V.41.N.2.1006.2003.doi:10.1029/2002RG000115
\bibitem{8}	Huder L.,Gillet N.,Finlay C.C.,Hammer M.D.,Tchoungui H.COV-OBS/
x2: 180 years of geomagnetic field evolution from ground-based and satellite observation //
 Earth,Planets and Space.72:160.2020.https://doi.org/10.1186/s40623-020-01194-2
\bibitem{9}	Chulliat A.,Hulot G.,Newitt L.R.Magnetic flux expulsion from the core as a possible cause of the unusually
large acceleration of the north magnetic pole during the 1990s /
/ J.Geophys.Res.V.115.B07101.2010.doi:10.1029/2009JB007143
\bibitem{10}	Olsen N.,Mandea M.Will the Magnetic North Pole Move to Siberia? //
 EOS Transactions AGU.V.88.? 29.P.293.2007a.
\bibitem{11}	Mandea M.,Dormy E.Asymmetric behavior of magnetic dip poles //
 Earth Planets Space.V.55.P.153-157.2003.
\bibitem{12}	Livermore P.,Hollerbach R.,Finlay C.An accelerating high-latitude jet in Earth's core.//
 Nature Geosci.V.10.P.62-68.2017.doi:10.1038/ngeo2859



\bibitem{CGN} A. Chulliat, Hulot Gauthier, L. R. Newitt. Magnetic flux expulsion from the core as a possible
cause of the unusually large acceleration of the north magnetic pole during the 1990s. Journal
of Geophysical Research: Solid Earth, American Geophysical Union, 2010, 115 (B7), pp.B07101.
ff10.1029/2009JB007143ff. ffinsu-01288849f.

\end{thebibliography}
\end{document}